\documentstyle[aps,preprint]{revtex}

\newcommand{\ben}{\begin{equation}}
\newcommand{\een}{\end{equation}}

\newcommand{\bena}{\begin{eqnarray}}
\newcommand{\eena}{\end{eqnarray}}
\newcommand{\mpr}{\mbox{$m^{\prime}$}}
\newcommand{\delU}{\mbox{$\Delta U$}}

\begin{document}
\draft

\title{
Escape-Rate Crossover between Quantum and Classical Regimes in
Molecular Magnets: A Diagonalization Approach
}

\author{
Gwang-Hee Kim$^1$ and  E. M. Chudnovsky$^2$
}

\address{\it{${}^{1}$Department of Physics, Sejong University,
Seoul 143-747, Republic of Korea \\
${}^{2}$Department of Physics and Astronomy, City University of New York-Lehman College, 
Bedford Park Boulevard West, Bronx, New York 10468-1589}}
\date{Received \hspace*{10pc}}
\maketitle

\thispagestyle{empty}

\begin{abstract}
We have studied numerically the quantum-classical crossover in 
the escape-rate for an uniaxial spin system with an arbitrarily directed field. 
Using the simple quantum transition-state theory,  we have obtained the 
boundary separating the first- and the second-order crossover and the escape-rate
in the presence of the transverse
and longitudinal field. The results apply to the molecular
nanomagnet, ${\rm Mn_{12}}$.
\\
\end{abstract}

We consider an easy-axis ferromagnetic nanoparticle, or a molecular cluster, that has metastable or degenerate classical spin states. The direction of the magnetization may change due to two mechanisms. At sufficiently high temperature the rate of the magnetization
reversal 
$\Gamma$ obeys the Arrhenius law, $\Gamma \sim \exp(-\delU/k_B T)$,
with $\delU$ being the height of the energy barrier. At a temperature low enough to
ignore the thermal activation, quantum tunneling comes into play with
$\Gamma \sim \exp(-\delU/\hbar \omega)$ where $\omega$ is 
some temperature-independent frequency related to the shape of the metastable potential well. 
The crossover between thermal and quantum regimes has 
been intensively studied in nanospin systems\cite{chu92,chu,gar,gmc,lmp,kim}.

This issue was first raised by
Chudnovsky and Garanin,\cite{chu} who observed that 
the crossover  in the spin Hamiltonian ${\cal{H}}=-D S^{2}_{z}-H_x S_x$ 
becomes sharp (first order)
for $ h_x (\equiv H_x/(2 DS)) <0.25$ and smooth (second order) for $0.25 \leq h_x  <1$. In the exponential approximation, when only the
transition exponent is concerned, the first- (second-) order crossover of the escpe-rate is
characterized by the discontinuity (continuity) of $d\Gamma(T)/dT$ at the
crossover temperature, $T_0$.
Subsequent calculations\cite{gmc} rendered the boundary between the first- and
the second-order crossover for the uniaxial model with a transverse and longitudinal field.
It was also pointed out \cite{chu92,chu,gmc} that the nonanalyticity of the rate for 
the first-order crossover disappears when quantum corrections to the exponential 
approximation are computed. The purpose of this Letter is to investigate 
the ``sharpness" of the first-order crossover. This question is especially 
important in the light of a recent experimental evidence of the first-order 
crossover in ${\rm{Mn}_{12}}$ \cite{ken}.

We shall focus on the crossover in the molecular magnet 
${\rm{Mn}_{12}}$ ($S=10$) \cite{fri} when the external magnetic field has both transverse and longitudinal components.
In order to calculate the splitting of the excited states, we shall
perform a numerical diagonaliztion of the Hamiltonian. Using numerical results,  we then obtain the group of levels which make the dominant contribution to the thermally
assisted tunneling. The full problem of the escape rate will be solved by mapping the spin problem onto a particle one \cite{zas}.
Summing contributions of all excited levels with account of quantum
corrections, we will show that the escape rate becomes analytic
for the first-order crossover, but changes sharply around $T_0$, in contrast with the second-order crossover.  
We also obtain the boundary between the two types of the crossover 
and find that the first-order regime is greatly suppresed by the longitudinal field in accordance with experiment \cite{ken}.

According to the simple quantum transition-state theory,\cite{wei} the escape-rate
in the temperature range $T \ll \delU$ is given by
\bena
\Gamma(T)={1 \over Z_0} \sum_m  \Gamma_m
\exp \left(-{E_m -U_{\rm min} \over k_B T} \right),
\label{rate}
\eena
where $\Gamma_m=\omega (E_m)  W(E_m)/(2 \pi)$, 
$\omega (E_m)$ is the frequency of oscillations at the energy $E_m$,
$W(E_m)$ quantum transition probabilities, and $Z_0$ the partition function in the
well. It is evident from Eq.  (\ref{rate}) that the nonanalyticity of the rate is not expected  around $T_0$ for any escape process, even though the crossover can be sharp because of the exponential dependences of $\Gamma_m$ and thermal populations on the parameters. 
The rate for the first-order crossover, $\Gamma_I (T)$ is not analytic
in the exponential approximation, when the escape rate is approximately given by the dominant term in the summation. However, the summation over all energy levels in
Eq. (\ref{rate}) smoothens this non-analyticity.

The model with an arbitrarily directed magnetic field
is described by the Hamiltonian
\bena
{\cal H}=-D S^{2}_{z}-H_z S_z -H_x S_x,
\label{hami}
\eena
The zero field Hamiltonian has uniaxial symmetry with easy axis along $z$ and
hard plane, $xy$. $H_z$ is  the longitudinal field
which affects the height of the energy barrier, and $H_x$ is the  transverse  field which 
is responsible for  quantum tunneling as well as for the reduction of  the barrier. 
In the first approximation, this spin model describes the magnetic molecule ${\rm Mn_{12}}$.

Within the thermally assisted model, the magnetization reversal occurs by quantum
tunneling from thermally excited magnetic levels at magnetic fields  which bring into resonance the
levels $m$ and $m^{\prime}$ 
belonging to different potential wells.
Denoting $m$ to be the escape level from the metastable well, the resonance
condition is that the levels $m$ and $m^{\prime}$ have the same energy when
$H_z = k D$ where $m^{\prime}=-m-k$ and $k$ is the bias index. 
The escape rate from any level is proportional to the product of the
probability of the thermal occupation of that level and the probability of
quantum tunneling from the level. In this respect, the dominat level
for a given temperature is determined by the function\cite{gmc}
\bena
f(m)= { \pi  (\Delta E_{m m^{\prime}})^2 \over 2  \omega (E_m)}
\exp \left(-{E_m -U_{\rm min} \over k_B T} \right),
\label{domi}
\eena
where it is assumed that the sum of the linewidths of the $m$-th and $m^{\prime}$-th levels
substantially exceeds the level spacing 
$\omega_{m^{\prime}}(=E_{m^{\prime}+1}-E_{m^{\prime}})$. Here
$E_m(=-D m^2 -H_z m)$ is the energy level of the spin system without transverse field,
and $\Delta E_{m m^{\prime}}$ is the splitting of the pair of in-resonance levels $m$ 
and $m^{\prime}$ on the opposite side of the anisotropy barrier.
It is seen from Eq. (\ref{domi}) that,
since  the escape rate  decreases exponentially with decreasing temperature,
larger longitudinal fields are necessary at lower temperature to produce an
observable tunneling rate.

Now, let us calculate the dominant level $m_d$ which maximizes $f(m)$. In order to
do that, we first need to find the range of $m$ in the metastable well for
a given transverse and longitudinal field, i.e., $ -S \leq m \leq m_t \leq 0$, when
$m_t$ is the level which is near the top of the barrier and inside the metastable
well. For $H_x=0$ and $H_z =k D$, simple analysis shows that $m^{0}_{t}=-[k/2]-1$
where $[x]$ gives the integer part of $x$. Since the height of barrier decreases with
increasing $H_x$, one expects $m_t  <  m^{0}_{t}$. To find the value of $m_t$ we express
Eq. (\ref{hami}) in the spherical coordinate and study the energy in the easy plane given
by
\bena
E(\theta, \phi=0)= -D S^2 (\cos^2 \theta+ 2 h_x \sin \theta+ 2 h_z \cos \theta),
\label{ene}
\eena
where $h_{x,z}= H_{x,z}/(2 D S)$. Writing the height of the barrier as $\delU\equiv DS^2 (\Delta u)$,
$m_t$ is determined by the relation
\bena
m_t = -\left[ {k \over 2}+ S \sqrt{(1-h_z)^2 -\Delta u} \right]  -1,
\eena
where
it is noted that, since $\Delta u=(1-h_z)^2$ in the absence of the transverse field, we obtain
$m_t=m^{0}_{t}$. Also, $m_t=-S \sqrt{h_x (2-h_x)}$ at $h_z=0$.
Numerical calculation of $\Delta u$ of Eq. (\ref{ene}) leads
to the results for $m_t (h_x, h_z)$ shown in Fig. \ref{figmt}.

Next, we consider the frequency of the real-time oscillations,
$\omega(E_m)$ at the energy $E_m$.
This quantity cannot be calculated with the use of the energy (\ref{ene})
in the spherical
coordinate system in which the physical quantity which is equivalent to
the mass of the system in a one dimensional case is unknown.
Thus, for the mapping of the spin problem
onto a particle one, the corresponding energy-dependent frequency is of the form
\bena
\omega(E)=2 \pi \left( {1 \over \sqrt{D}} \int^{x_2(E)}_{x_1(E)}
 {dx \over \sqrt{E- U(x)}} \right)^{-1}
\label{ome}
\eena
where $x_{1,2}$ are turning points in the particle potential $U(x)$ for a given energy $E$.
Introducing the parameter\cite{chu} $p=(U_{\rm sad}-E_m)/\Delta U$, the specific form
of $p$ at a small value of $h_x$ becomes,
\bena
p=\left( {{m \over S} + h_z \over 1-h_z } \right)^2,
\eena
where $p=0$ at the top of the barrier. This gives $m/S=-h_z$, i.e.,
$m=-k/2$ which is related to $m_t$ discussed previously.
After some trivial manipulation of Eq. (\ref{ome}), we obtain the dependence of
$\omega$ on $h_x$, $h_z$, and $m$. As a result, the frequency of oscillation in $f(m)$
can be numerically deduced by taking $h_x \rightarrow 0$.
In this limit  the frequency also can be computed for the spin model having
the energy levels $E_m=-Dm^2-H_z m$ as the inversed density of states, i.e.,
the energy difference between neighboring levels at energy $E$. This is
simply given by $\omega_m \simeq -D(2 m+k)$ for $S \gg 1$.
Now, in order to calculate the level splitting $\Delta E_{mm^{\prime}}$, we first
consider the formula of the perturbation theory\cite{gar}
\bena
\Delta E_{m m^{\prime}}& =&{2 DS^{m^{\prime}-m} \over [(\mpr-m-1)!]^2 } 
\nonumber \\
&&\times \left[ { (S+\mpr)! (S-m)! \over  (S-\mpr)! (S+m)! }  \right]^{1/2} h^{m^{\prime}-m}_{x}.
\label{pert}
\eena
This formula is compared with the results from the direct numerical
diagonalization.  As is illustrated in Table \ref{tabpd}, there is a disagreement between
them for levels with $m \lesssim -6$. However, noting that $m_t(h_x=0, h_z=0.1)=-2$
is shifted to $m_t(h_x=0.05, h_z=0.1)=-6$, the level $m=-6$ or $ -7$
is important to study the type of the crossover and
the escape-rate around $T_0$. Accordingly, we will perform the
numerical diagonalization for the level splitting of $f(m)$.

Now, we are in a position to calculate the dominant level $m_d$
which is determined by the maximal value of the function (\ref{domi})
for a given temperature. Within the thermally assisted tunneling model
the system tunnels through the levels between the bottom and
the top. In this process $m_d(T)$ behaves in two different ways.
One way is that $m_d$ changes continuously from $-S$ to $m_t$,
whereas the other way is that it performs some discontinuous jump.
We call the former - the second-order crossover and the latter - the first-order
crossover. As is shown in Fig. \ref{figjump}, for the resonant field 
$h_z=0.1$ we have  the first-order crossover for $h_x=0.05, 0.1$
and  the second-order crossover for $h_x=0.15$. Also,
for $h_z=0.4$ ( $H_z \simeq 3.28$ Tesla in ${\rm Mn_{12}}$ ),
the abrupt shift occurs at $h_x=0.04$
and the corresponding dominant level changes  by 2 ($m_d=-10$ and
$m_d=-8$) in the range of temperature ($\sim$ 0.1 K -- $\sim$ 1 K ).
Strikingly, this feature is observed in a recent experiment,\cite{ken}
in which the step positions shift abruptly at low temperature and
high magnetic field.
Employing these schemes in the whole range of $h_z$, we obtain
the phase boundary for the values of the transverse field, which 
is shown by the symbols in Fig. \ref{figphase}.
In the quasiclassical method, the order of the quantum-classical
escape-rate crossover was determined by the sign of the
coefficient in the expansion of the imaginary-time action near the top of 
the barrier\cite{gmc}. In the perturbation method, the behavior of the result
(\ref{pert}) is inserted into the calculation of the dominant level
$m_d$ in the metastable well, whose behavior determines the type of the order.
Using this method, the first-order crossover is found to be suppresed as compared
to the quasiclassical method. This was noticed in Ref. \cite{gar}.
However, as discussed previously, the corresponding level splitting
is not quite correct, especially, near the top of the barrier, in which
range the perturbation fails. The correct calculation based on the
diagonalization method shows that the first-order regime is even more
suppresed as compared to the perturbative results.
For example, in the unbiased case, the phase boundary becomes at $h_x=$ 0.114, 0.139, and 0.25
for the diagonalization, perturbation, and quasiclassical method, respectively.
In Fig. \ref{figphase},  there  is no data point beyond $h_z = 0.8$.
The reason is that $m_t=-9$ in this region and thereby it is meaningless
to ask whether the shift is  continuous or discontinuous in this region.

The values of the crossover temperature $T^{(c)}_{0}$ at the
phase boundary between first- and second-order crossover,
which have been obtained by the diagonalization method described
above, are shown in Fig. \ref{figcross}. Comparing this with two other methods,
the result based on the diagonalization method
is the smallest in the whole range of the bias field.
For example, in the unbiased case, we have $T^{(c)}_{0}/(DS)=$ 0.117, 0.124, and
0.137 for the diagonalization, perturbation, and quasiclassical method, respectively.

Now, for the relaxation at resonance
we will present the results of numerical calculation
for the escape rate, $\Gamma(T)$.  
As is
shown in Fig. \ref{figrate} (as, e.g., for $h_z=0.1$), the escape rate
in the first-order  region ($h_x=0.05$ or 0.1) changes sharply, as shows the comparison
with the escape rate in the second-order one ($h_x=0.15$ or 0.2). 
Also, for $h_z=0.4$ we can clearly distinguish the behavior of the rates in two
different regimes, e.g., $h_x=0.04$ and 0.1, and
its trend continues in
the whole range of the bias field. The origin of these behaviors is that,
as described above, $m_d$ changes discontinuously around $T_0$ for the first-order
crossover, while it does continuously for the second-order one. In other words,
since the rate (\ref{rate}) is sensitive to the change of $m$, the abrupt jump in
the first-order crossover induces sharp increase, e.g., 
$\Gamma(h_x=0.05, h_z=0.1)/\Gamma(h_x=0.2, h_z=0.1) 
\simeq 10^6$ at  $T/(DS) \simeq 0.12$, and 
$\Gamma(h_x=0.04, h_z=0.4)/\Gamma(h_x=0.1, h_z=0.4) 
\simeq 10^5$ at  $T/(DS) \simeq 0.11$.

For the uniaxial spin model considered in this paper, both type of crossover can
be realized in the molecular magnet ${\rm Mn_{12}}$, and the situation
can be controlled by the longitudinal and transverse field. For ${\rm Mn_{12}}$,
$D \simeq 0.55 $K, and
$\delU=DS^2 \simeq 55 $K at $h_x=h_z=0$\cite{bar,mir}. The critical field
in the $x$- or $z$-direction  is $H_{xc}=H_{zc}=2 DS/(g \mu_B) \simeq 8.2$T, and
the longitudinal field for the resonance is $H_0 = D/(g \mu_B) \simeq 0.41$T.
At these fields, the magnetic relaxation in ${\rm Mn_{12}}$ can occur on measurement
time scales, and gives rise to the different behavior of the dominant level and the rate
in magnetization depending on $H_x$ and $H_z$. Furthermore, in the unbiased case
the first-order crossover can be observed in the field
range $ 0<H_x < 0.93$ Tesla and the crossover region occurs at the
temperature range $\sim 0.1~ {\rm K} < T < \sim 1 ~ {\rm K}$. This field range decreases with increasing
the bias field.
Even though we presented the results of $m_d$ only for $h_z=0.1$ and
0.4  in Fig. \ref{figjump},
we have found that the levels which, at a certain temperature, change by more than 1 are
$m=7$, 8 or 9 depending on the value of the resonant field.

In conclusion, we have studied the quantum-classical crossover of the escape-rate of
a uniaxial spin model with an arbitrarily directed field. Employing the diagonalization
method, we have obtained the dominant level for the thermally assisted
tunneling and  dependence of the
escape rate on temperature. In comparison to the previously studied models, 
the first-order region is greatly suppressd,
but still observable in molecular magnets.
It is also found that the first-order crossover is fairly sharp while the second-order the crossover is smooth.
This is found to be  strongly related with the discontinuous (continuous) jump of $m_d$
in lower (higher) field $h_x$. These results have been applied to the high-spin molecule, 
${\rm Mn_{12}}$. They are also relevant to the study of  nanoparticles. 

G.-H. K. is grateful to A. Garg for  useful discussions.
E. C. acknowledges support from the NSF Grant DMR-9978882.
G.-H. K. was  supported 
by grant No. 1999-1-114-002-5 from the Interdisciplinary
Research Program of the KOSEF.\\

\begin{table}
\caption{
Comparison of the splitting (\protect\ref{pert}) with the one
from the numerical diagonalization for $S=10$,
$h_x=0.05$, and $h_z=0.1$.}
\label{tabpd}
\end{table}

\begin{center}
\begin{tabular}{|c|c|c|}  \hline \hline
 $m$ & pert. & diag. \\
\hline
-6 & $0.156$ & $8.53 \times 10^{-2}$ \\
-7 & $1.01 \times 10^{-3}$ & $7.38 \times 10^{-4}$ \\
-8 & $2.72 \times 10^{-6}$ & $2.26 \times 10^{-6}$ \\
-9 & $3.14 \times 10^{-9}$ & $2.78 \times 10^{-9}$ \\
-10 & $1.40 \times 10^{-12}$ & $1.28 \times 10^{-12}$ \\
\hline \hline
\end{tabular}
\end{center}

\begin{figure}
\caption{
$m_t /S$ vs. $h_x$ for a given value of $h_z$.
The range of $h_z$ is $0 \sim 0.9$.
}
 \label{figmt}
\end{figure}

\begin{figure}
\caption{$m_d$ vs. $\bar{T} (\equiv T/(DS))$
for $h_z=0.1$, where $h_x=$ 0.05 (a), 0.1 (b), and 0.15 (c).
Inset: $h_z=0.4$, where $h_x=0.04$ (a), 0.06 (b) and 0.1 (c).
}
\label{figjump}
\end{figure}

\begin{figure}
\caption{
Phase boundary between the first- and the second-order crossover
obtained by the quasiclassical (b), the perturbative (c) and
the diagonalization method (d). 
The critical field\protect\cite{sto} ($h^{2/3}_{xc}+h^{2/3}_{xc}=1$) is 
represented in (a).
}
\label{figphase}
\end{figure}

\begin{figure}
\caption{
Crossover temperature $\bar{T}^{(c)}_{0}( \equiv T^{(c)}_{0}/(DS) )$ at the phase boundary between
first- and second-order crossover, based on 
the quasiclassical (a), the perturbative (b) and the
diagonalization method (c)
}
\label{figcross}
\end{figure}

\begin{figure}
\caption{ $\bar{\Gamma} ( \equiv \Gamma(T)/\Gamma(0))$ 
vs. $\bar{T} ( \equiv T/(DS))$ for $h_z=0.1$ ,
where $h_x=$  0.05 (a), 0.1 (b), 0.15 (c), and  0.2 (d).
Inset: $h_z=0.4$, where $h_x=$ 0.04 (a), 0.06 (b) and 0.1 (c).
}
\label{figrate}
\end{figure}

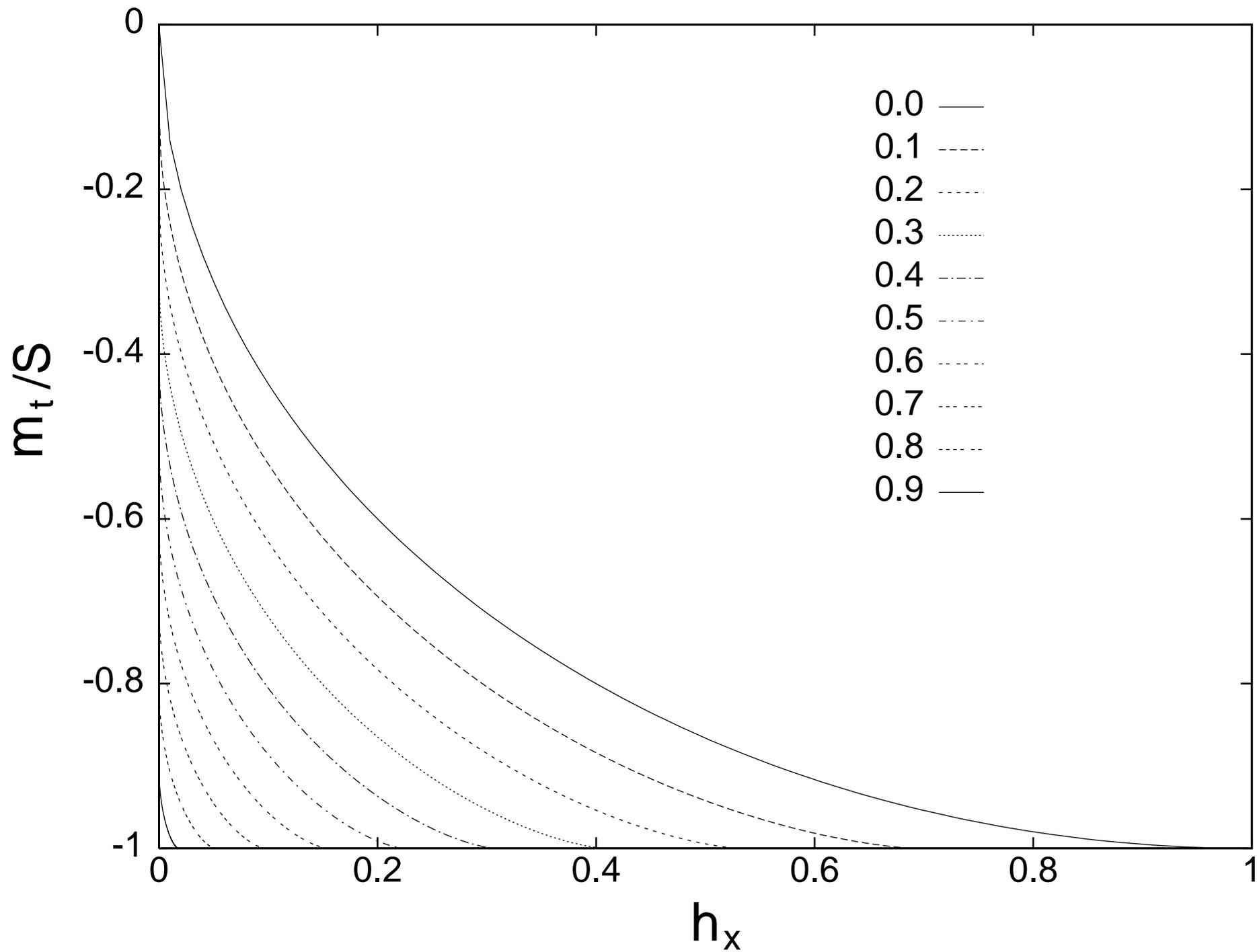
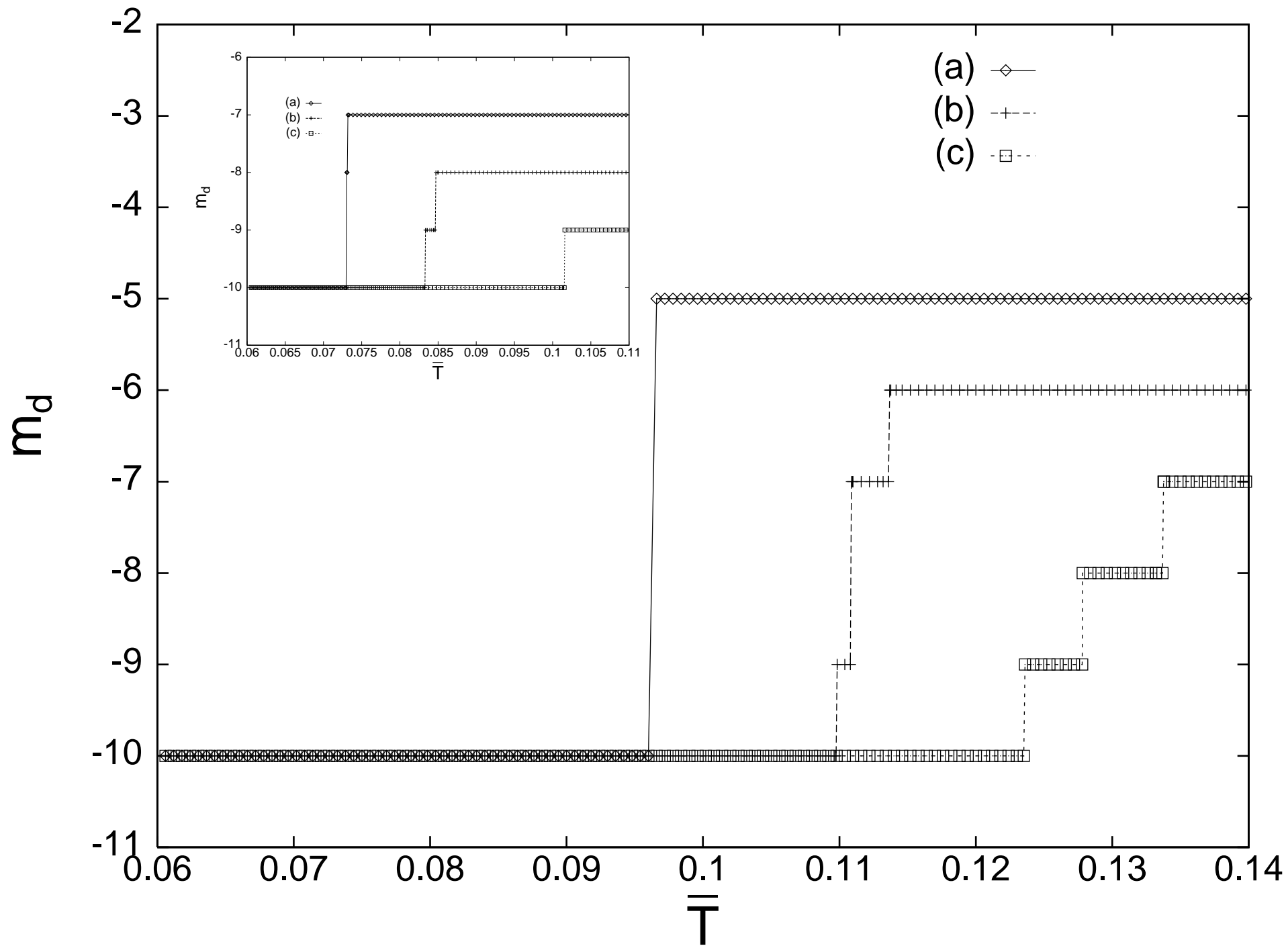
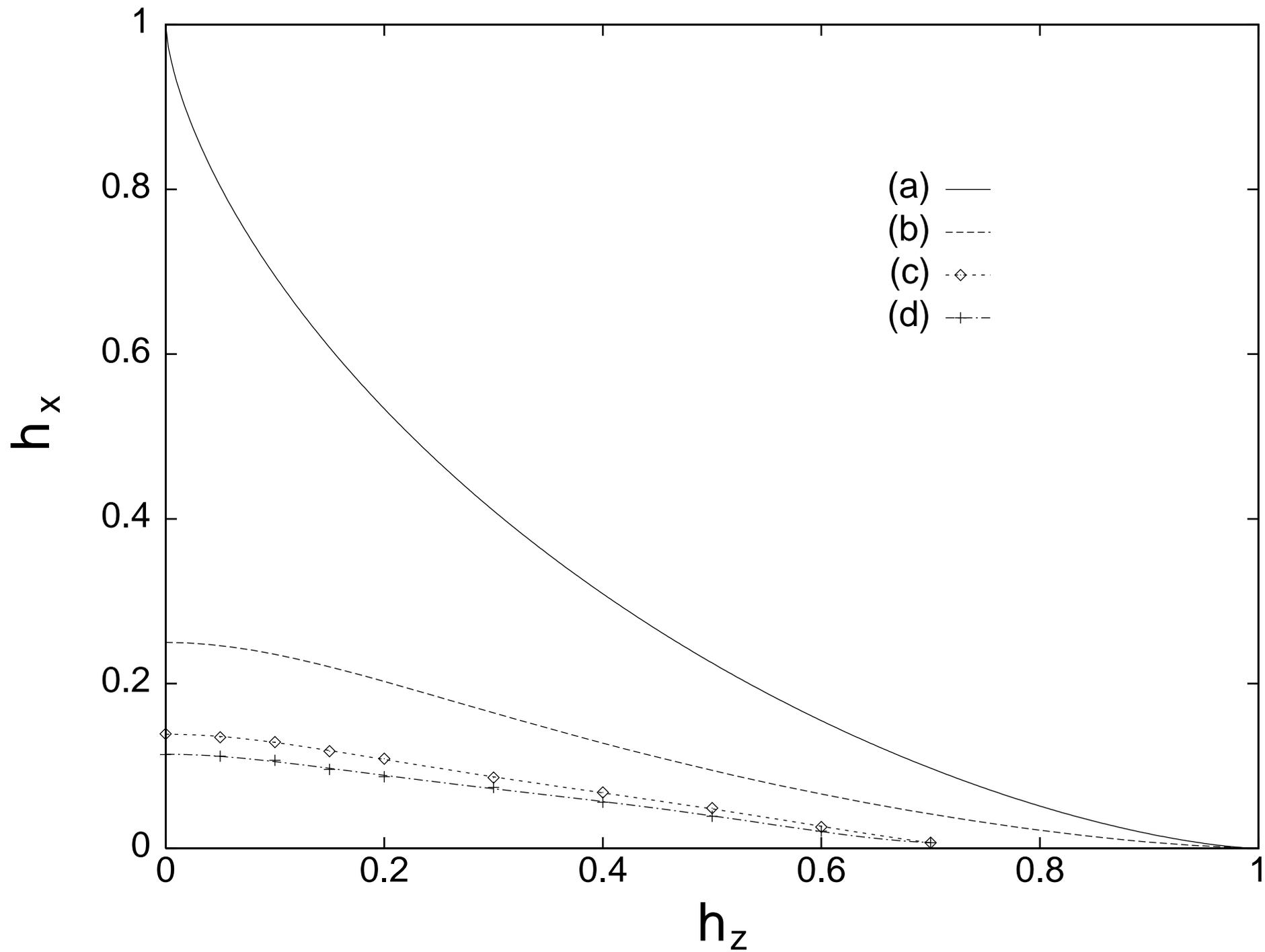
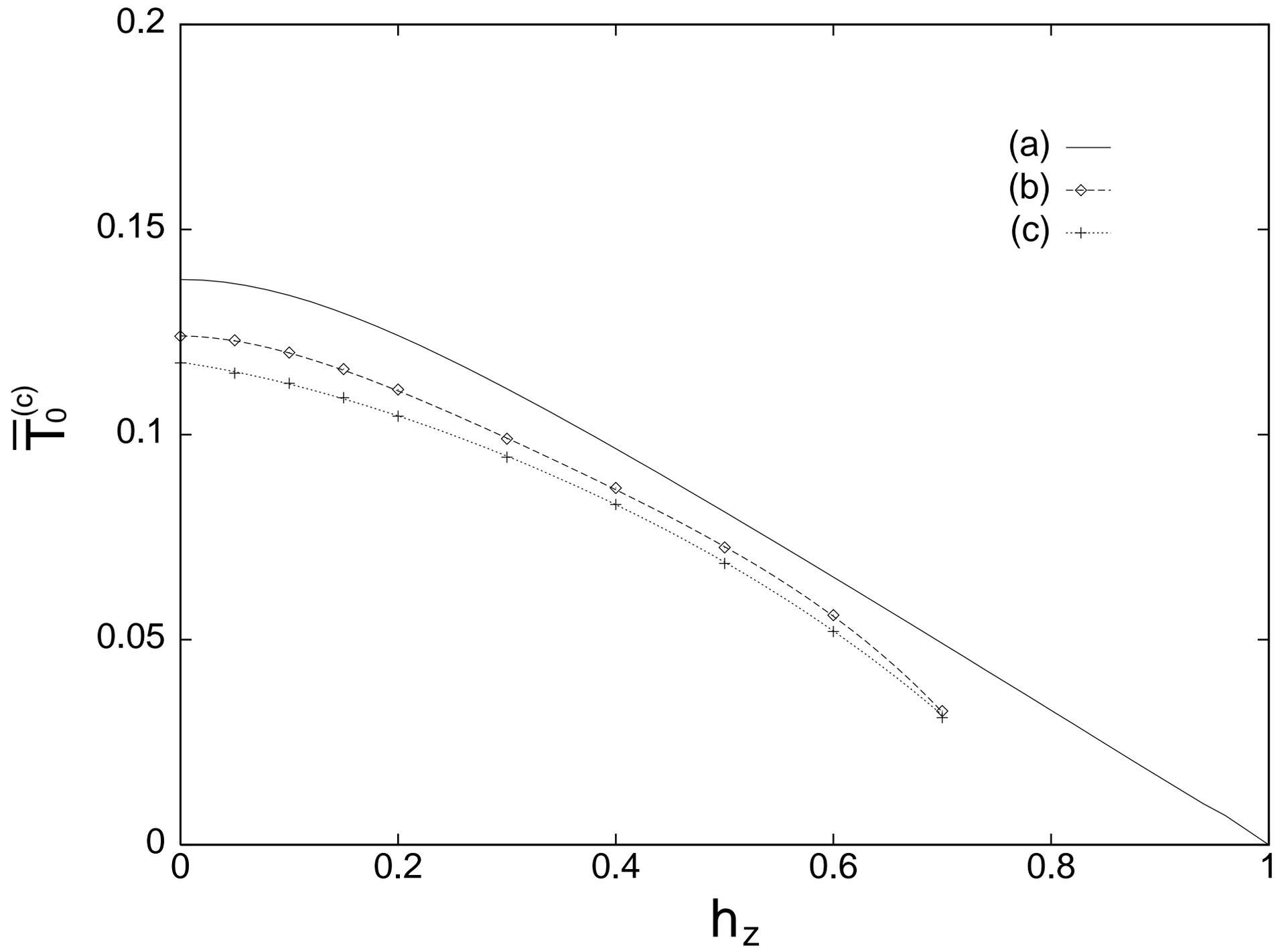
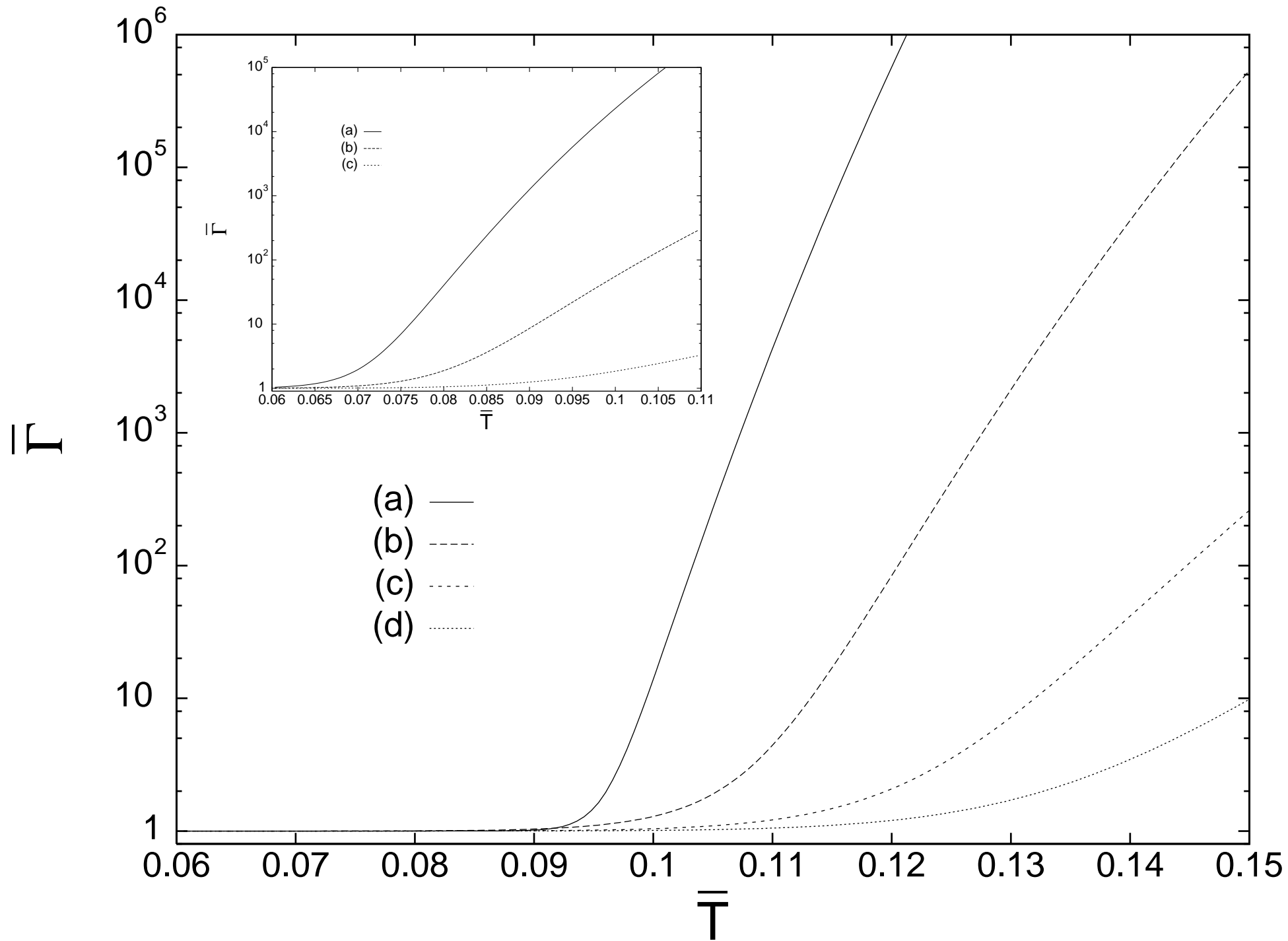
\end{document}